\documentclass{article-cfip}

\usepackage{amsmath}
\usepackage{subfig}
\usepackage{graphicx}
\usepackage{array}
\usepackage[utf8]{inputenc}
\usepackage[pdftex,unicode]{hyperref}

\begin{document}

\title[ICMN et Dimensionnement de Messages]{Dimensionnement des messages dans un réseau \\ mobile opportuniste}

\author{John Whitbeck\fup{*,**} \andauthor Vania Conan\fup{*} \andauthor Marcelo Dias de Amorim\fup{**}}
\address{
  \fup{*}Thalès Communications \\[6pt]
  \fup{**}LIP6/CNRS - UPMC Paris Universitas
}
\proceedings{CFIP'09}{-}

\resume{ Comprendre la capacité des réseaux mobiles à connections
  intermittentes (ICMN - Intermittently Connected Mobile Networks) est
  important pour bien répondre aux divers besoins d'interactivité et
  de bande passante des applications mobiles. Un problème pratique
  concerne la transformation des messages d'une application en paquets
  prêts à être transmis à travers un ICMN. Dans ce papier, nous
  proposons un nouveau modèle Markovien de graphes aléatoires
  temporels et montrons, à la fois analytiquement et en rejouant une
  trace de connectivité obtenue lors d'une randonnée roller, que la
  taille des messages envoyés dans un ICMN a un impact décisif sur
  leur taux de livraison. Cependant, nous montrons également que ce
  gain de fiabilité n'apparaît que sous l'effet de fortes contraintes
  sur le délai maximal toléré. Ainsi, il faut ajuster la taille des
  messages en fonction à la fois des besoins des applications et de la
  dynamique de la topologie du réseau pour améliorer les performances
  du routage.  }

\abstract{ Understanding transport capacity in intermittently
  connected mobile networks (ICMN) is crucial since different
  applications have different interactivity and bandwidth
  requirements. One practical issue is how to transform an
  application's messages into packets suitable for transport over an
  ICMN. In this paper, we propose a new Markovian model for random
  temporal graphs and show, both analytically and by replaying a real
  life trace obtained in a rollerblading tour, that the size of the
  messages sent over an ICMN has a decisive impact on their delivery
  ratio. A given message could therefore be broken down into smaller
  packets to increase reliability. However, we also show that this
  gain in reliability only appears under tight constraints on the
  maximum delay tolerated. Mobile application designers should
  therefore balance message size against both application requirements
  and network topology dynamics to improve performance.  }

\motscles{Réseaux tolérants aux délais (DTN), Réseaux mobiles à connections intermittentes (ICMN), Taille des messages, Applications Mobiles}


\maketitlepage

\section{Introduction}
\label{sec:introduction}

La prolifération d'appareils mobiles, tels que les téléphones
portables, les consoles de jeux ou lecteurs multimédias, dotés de
multiples interfaces sans fil, offre de nombreuses nouvelles
opportunités de communications ad-hoc. Les processus sociaux qui
mettent ces appareils en contact donnent naissance à des réseaux
mobiles à connectivité intermittente (ICMN). A cause de la grande
mobilité des n{\oe}uds et de l'absence fréquente de connectivité de bout
en bout, le transport des messages dans de tels réseaux est
typiquement assuré en stockant entièrement le message à chaque saut
avant de le transmettre au suivant (\emph{store-and-forward}).

Toutefois, un telle approche esquive certains problèmes plus
concrets. Différentes applications mobiles ont des besoins différents
d'interactivité et de bande passante, et doivent gérer la pénurie de
ressources dans un ICMN. Elles doivent toutes notamment diviser leurs
messages en paquets avant de les transmettre sur le réseau. Ceci
nécessite non seulement une évaluation de l'impact que la taille de
ces paquets aura sur le taux de livraison et le délai, mais également
une compréhension de la relation entre taille optimale des paquets et
dynamique du réseau.

Dans ce papier, nous examinons, analytiquement et expérimentalement,
l'influence de la taille des paquets sur le taux de livraison, sous la
contrainte d'un délai maximum toléré par l'application. Nous
présentons trois contributions. Tout d'abord, nous proposons un
nouveau modèle de graphe aléatoire temporel fondé sur des chaînes de
Markov qui capture la corrélation des états des liens lors de pas de
temps successifs et permet l'étude de la relation entre stabilité des
liens et taux de livraison. A partir de ce modèle, nous montrons que
le taux de livraison augmente pour des paquets plus petits, mais que
le gain ainsi obtenu est borné et n'est significatif qu'en présence de
fortes contraintes sur le délai. Finalement, nous comparons nos
résultats théoriques avec ceux obtenus à partir d'une trace de
connectivité collectée lors d'une randonnée roller.

L'étude des graphes aléatoires temporels en est encore à ses
débuts. Les travaux précédents se focalisaient sur des graphes avec un
nombre croissant de sommets et d'arêtes~\cite{GOT}. Toutefois, ceux-ci
ne rendent pas compte de la mobilité des n{\oe}uds et/ou de l'instabilité
des liens. Plus récemment, des suites de graphes aléatoires uniformes
ont été proposés pour modéliser des graphes aléatoires
temporels~\cite{chaintreau_diam}. Cependant, de tels modèles ne
capturent pas la corrélation entre états successifs du graphe de connectivité.

Le modèle de graphe temporel Markovien que nous proposons comble ce
vide. En étudiant la propagation épidemique de messages sur ce modèle,
nous calculons le meilleur taux de livraison possible pour tout
protocole de routage de type ``store-and-forward''. Nous montrons que
ceci amène une meilleure compréhension des interactions entre la
mobilité des n{\oe}uds, les contraintes de délais et la taille des
paquets, et comment ceux-ci affectent le taux de livraison.

Dans la prochaine section, nous décrivons notre approche et nos
hypothèses. Dans la section~\ref{sec:theory}, nous décrivons notre
modèle de graphe temporel Markovien. Nous calculons ensuite le taux de
livraison pour le routage épidemique dans la section~\ref{sec:epidemic}
et montrons l'impact de la taille des paquets sur le taux de
livraison. Nous validons ensuite ces intuitions théoriques sur une
trace de connectivité expérimentale, l'expérience Rollernet, dans la
section~\ref{sec:experimental}.

\section{Préliminaires}
\label{sec:motivation}

\subsection{Travaux connexes}
\label{subsec:related}
La topologie d'un réseau d'appareils mobiles évolue au cours
du temps au gré des apparitions et chutes de liens. La succession des
instantanés de ce graphe de connectivité évolutif donne un
\emph{graphe temporel}, une suite indexée par le temps de graphes de
connectivité statiques. L'ajout de cette composante temporelle se
traduit par de nouvelles métriques comme, par exemple, l'existence de
chemins spatio-temporels entre deux sommets même en l'absence, à tout
instant, de chemin de bout en bout entre eux. 
Puisque de tels graphes temporels apparaissent naturellement lors de
l'étude de traces de connectivité dans lesquelles les participants
sondent leur voisinage à intervalles réguliers, leur étude théorique
est importante pour la compréhension de la dynamique sous-jacente aux
réseaux mobiles.

Les résultats théoriques sur la connectivité, la dynamique et la
performance de protocoles de routage dans les réseaux ad-hoc (au sens
large) s'obtiennent typiquement ou bien à partir de simulations sur des
modèles de mobilité~\cite{leboudec05,lenders08}, ou bien à partir de
modèles à base de graphes aléatoires
temporels~\cite{pellegrini07}. Les premiers disposent d'un modèle
physique plus réaliste, alors que les seconds sont plus simples et
permettent le calcul explicite de grandeurs topologiques (distribution
du degré des n{\oe}uds ou de la longueur des chemins, par exemple) et
de métriques de performance (taux de livraison, délai, par exemple).
Ces approches doivent toutes deux être confrontées aux données
expérimentales issues de traces de connectivité réelles, où
l'attention s'est focalisée sur la distribution des temps
d'inter-contact. Lorsque les dynamiques sociales sous-jacentes sont
fortes, cette distribution suit typiquement une loi de
puissance~\cite{chaintreau_mobility}. Cependant, dans d'autres
scénarios, elle peut suivre une loi exponentielle~\cite{lenders08}. De
façon intéressante, il suffit de retirer aux modèles de mobilité
classiques leurs frontières pour passer d'une loi exponentielle à une
loi de puissance~\cite{cai07}. Les approches fondées sur des graphes
aléatoires, y compris la nôtre, présentent une distribution
exponentielle (ou géométrique) des temps d'inter-contact.

La modélisation de réseaux mobiles dynamiques à partir de graphes
aléatoires temporels est un domaine assez nouveau. De simples suites
de graphes aléatoires réguliers peuvent être utilisées pour analyser
le diamètre de réseaux mobiles
opportunistes~\cite{chaintreau_diam}. La notion de \emph{d'emergence
  de connectivité spatio-temporelle} a été étudiée mais celle-ci perd
toute information sur l'ordre dans lequel ces opportunités de contact
sont apparues~\cite{pellegrini07}. Dans ce papier, nous améliorons ces
travaux en capturant la forte corrélation qui existe entre les graphes
de connectivité aux instants $t$ et $t+1$.
Pour cela, nous proposons un modèle Markovien de graphes
aléatoires temporels dans la section~\ref{sec:theory}. Des résultats
asymptotiques pour l'inondation de ce genre de graphes, lorsque le
nombre de n{\oe}uds tend vers l'infini, ont été
obtenus~\cite{Clementi08}. Nos résultats, en revanche, concernent des
paires source/destination, un nombre de n{\oe}uds fini, ainsi que des
capacités et des tailles de message finies.

\subsection{Hypothèses}
\label{subsec:hypotheses}
\begin{table}[t]
  \centering
\caption{Notations utlisées dans ce papier.}
\begin{tabular}{cl}
  \textbf{Notation} & \textbf{Description} \\
  \hline
  $N$ & nombre de n{\oe}uds \\
  $\tau$ & pas de temps \\
  $\alpha$ & taille des paquets \\
  $d$ & délai maximum (en nombre de pas de temps)\\
  $r$ & durée de vie moyenne d'un lien \\
  $\lambda$ & fraction du temps pendant laquelle un lien est ``down'' \\
  \hline
\end{tabular}
  \label{param_desc}
\end{table}

Dans la suite de ce papier,
nous considérons des graphes temporels de $N$ n{\oe}uds mobiles qui
évoluent en temps discret. Le pas de temps $\tau$ est le plus petit
temps de contact ou d'inter-contact. Sur une trace réelle, $\tau$ sera
égal à la période d'échantillonage. La seule différence entre pas de
temps successifs est l'état des liens. Ces derniers peuvent apparaître
ou disparaître au début de chaque pas de temps, mais la topologie
reste ensuite statique jusqu'au pas de temps suivant.
Lorsqu'ils existent, tous les liens partagent le même débit $\phi$ et
peuvent donc transporter la même quantité $\phi \tau$ d'information
pendant un pas de temps. Nous appelons $\phi \tau$ la \emph{capacité
 du lien}. La taille des paquets vaut $\alpha \phi \tau$ où $\alpha$
peut être plus grande ou plus petite que $1$. Par abus de langage, nous
appelons $\alpha$ la taille des paquets. Par exemple, un paquet de
taille $2$ ($\alpha=2$) pourra uniquement traverser des liens qui
subsistent plus que $2$ pas de temps, alors qu'un paquet de taille
$0.5$ pourra effectuer deux sauts pendant chaque pas de temps. La
taille des paquets ainsi définie est proportionnelle à $\tau$.
De petites valeurs de $\tau$ indiquent que le temps caractéristique
d'évolution de la topologie du réseau est court, et donc que seules de
petites quantités d'information peuvent transiter sur un lien pendant
un pas de temps. De plus, nous supposons qu'une application mobile ne
peut tolérer qu'un certain délai maximum. Nous notons $d$ le nombre de
pas temps au-delà duquel on considère que la livraison d'un paquet à
échoué. Toutes ces notations sont résumées dans le
tableau~\ref{param_desc}.

\section{Graphes temporels Markoviens}
\label{sec:theory}
Dans cette section, nous introduisons une nouvelle famille de graphes
aléatoires temporels qui utilisent une chaîne de Markov discrète pour
modéliser la transition d'un graphe de connectivité à l'instant $t$ à
son successeur à $t+1$.

Considérons un réseau à $N$ n{\oe}uds. Chacun des $\frac{N(N-1)}{2}$ liens
évolue indépendamment et peut se trouver dans l'un des deux états
$\uparrow$ ou $\downarrow$. Plutôt que d'utiliser, comme dans un
graphe aléatoire classique, une probabilité $p$ d'être dans l'état
$\uparrow$, nous modélisons chaque lien par un chaîne de Markov à
deux états où $q_c$ (resp. $q_i$) est la probabilité que le lien reste
dans l'état $\uparrow$ (resp. $\downarrow$). L'évolution du graphe de
connectivité peut ainsi s'envisager comme une chaîne de Markov sur le
produit tensoriel des états de ses liens. Cependant, ses
$2^\frac{N(N-1)}{2}$ états font qu'elle est trop lourde à manipuler en
pratique.

A chaque pas de temps, chaque lien effectue une transition dans sa
chaîne de Markov.
Si $0 < q_i < 1$ et $0 < q_c < 1$, cette chaîne est positive,
récurrente et apériodique, et donc ergodique.
Dans la suite du papier, nous utiliserons les deux paramètres suivants~:
\begin{equation}
r = \frac{1}{1-q_c} \quad ; \quad \lambda = \frac{1-q_c}{1-q_i} .
\end{equation}
Les temps de contact ($T_c$) et d'inter-contact ($T_i$) suivent une
distribution géométrique et leurs espérances sont~:
\begin{equation}
E(T_c) = r \tau \quad ; \quad E(T_i) = \lambda r \tau .
\label{contact_inter_contact_times}
\end{equation}
Soit $\pi_\uparrow$ (resp. $\pi_\downarrow$) la probabilité
stationnaire d'être dans l'état $\uparrow$ (resp. $\downarrow$)~:
\begin{equation}
\pi_\uparrow = \frac{1}{1+\lambda} \quad ; \quad \pi_\downarrow = \frac{\lambda}{1+\lambda}.
\end{equation}

Ici $r$ est le nombre moyen de pas de temps qu'un lien passe dans
l'état $\uparrow$, tandis que $\lambda$ est la fraction du temps qu'un
lien passe dans l'état $\downarrow$. D'une certaine façon, $r$ mesure
la vitesse d'évolution de la topologie du réseau, tandis que $\lambda$
est lié à sa densité. La durée de vie moyenne d'un lien est, par
définition, $r \tau$, tandis que le degré moyen vaut
$\frac{N-1}{1+\lambda}$. Puisque nous considérons un temps discret, un
lien ne peut pas demeurer moins de $1$ pas de temps dans un état
donné. Ainsi $r \ge 1$ et $\lambda \ge \frac{1}{r}$.

\section{Propagation épidémique avec contrainte de délai}
\label{sec:epidemic}

Notre but ici est de calculer la probabilité de livraison d'un paquet
en utilisant du routage épidémique dans notre modèle de réseau
aléatoire Markovien. Pour réussir, le paquet doit atteindre sa
destination avant le délai maximal toléré. Le routage épidémique est
utile pour des raisons théoriques car son taux de livraison est aussi
celui du routage spatio-temporel optimal avec une unique copie du
paquet.

Nos paramètres sont le nombre de n{\oe}uds $N$, les paramètres $r$ et
$\lambda$ du modèle de lien, le délai maximum $d$ (en nombre de pas de
temps) et la taille du paquet $\alpha$. Pour des raisons de
simplicité, nous présenterons, dans un premier temps, le modèle pour
$\alpha=1$. Dans les sections~\ref{subsec:p_smaller}
et~\ref{subsec:p_greater}, nous décrirons respectivement comment
adapter le modèle aux cas où les paquets sont plus petits ($\alpha<1$)
ou plus grands ($\alpha>1$) que la capacité du lien. Finalement, nous
analyserons l'influence de chaque paramètre sur le taux de
livraison. En particulier, nous montrerons comment le taux de
livraison augmente pour de petites tailles de paquets, mais que le
gain ainsi obtenu est borné et n'est significatif que lorsque les
contraintes sur le délai sont très fortes.

\subsection{Paquets de taille égale à la capacité du lien \texorpdfstring{($\alpha = 1$)}{}}
%\subsection{Paquets de taille égale à la capacité du lien ($\alpha = 1$)}
\label{subsec:epi_model}

\begin{figure}[t]
  \centering
  \includegraphics{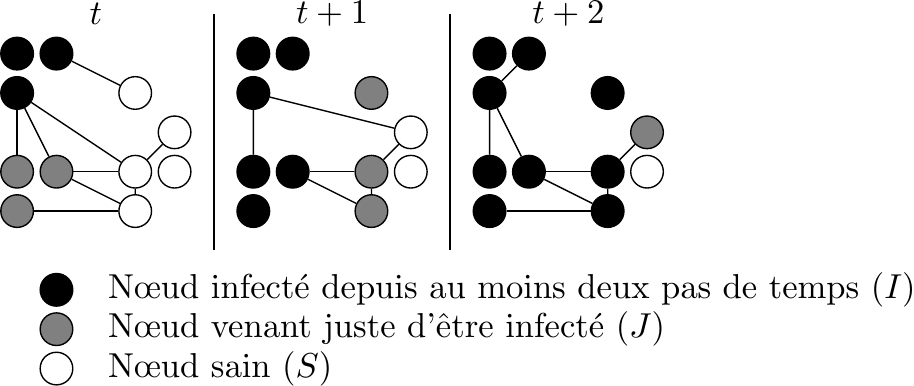}
  \caption{Propagation épidémique dans un réseau aléatoire Markovien.}
  \label{epifig}
\end{figure}

La source $a$ désire envoyer un paquet à la destination $b$ en
utilisant du routage épidémique. Dans la suite, on dira d'un n{\oe}ud
disposant d'une copie du paquet qu'il est \emph{infecté}. En un pas de
temps, un n{\oe}ud infecté peut uniquement infecter ses voisins
directs puisque la taille du paquet est $1$ et ne peut donc effectuer
qu'un seul saut par pas de temps. Cet infection à lieu au \emph{début}
du pas de temps. Soit $V$ l'ensemble des n{\oe}uds du réseau. Après $k$
pas de temps, les n{\oe}uds autres que $b$ se répartissent en l'une
des trois catégories suivantes~:
\begin{itemize}
\item ceux qui viennent \emph{juste} d'être infectés~: $J_k$~;
\item ceux qui sont infectés depuis au moins deux pas de temps~: $I_k$~;
\item ceux qui sont encore sains~: $S_k = V \setminus (I_k \cup J_k \cup \{b\})$.
\end{itemize}

Cette distinction est nécessaire pour déterminer qui est susceptible
de se faire infecter au pas de temps $k+1$. En effet, si un n{\oe}ud
appartient à $I_k$, alors tous ses voisins au pas de temps $k$ sont
dans $I_k \cup J_k$. Il peut uniquement infecter d'autres n{\oe}uds si un
lien vers un n{\oe}ud sain dans $S_k$ apparaît au pas de temps
$k+1$. Cependant, un n{\oe}ud dans $J_k$ peut avoir des voisins dans
$S_k$. Si ces liens se maintiennent au temps $k+1$, ces voisins
deviendraient infectés à leur tour (fig.~\ref{epifig}).

Dans ce papier, nous nous intéressons uniquement à la probabilité que
$b$ reçoive une copie $d$ pas de temps ou moins. Dans ce cas, la
seule information nécessaire pour caractériser l'état de l'épidémie
est le nombre $i$ et $j$ de n{\oe}uds dans les états $I_k$ et $J_k$
respectivement. Le taux de livraison peut être obtenu comme la
probabilité d'absorption de la chaîne de Markov décrite ci-dessous.

\vspace{2mm}\noindent\textbf{États.} Nous décrivons l'épidémie par une
chaîne de Markov sur les $2+\frac{N(N-1)}{2}$ états suivants~:
\begin{itemize}
\item $Init$~: l'état initial où seule l'origine $a$ est infectée (Cet
  état est transitoire)~;
\item $Succ$~: l'état où la destination $b$ a été infectée (Cet état
  est absorbant et indique que le routage a réussi)~;
\item les états $(i,j)$ pour $1 \le i \le N-1$ et $0 \le j \le   N-1-i$.
\end{itemize}

\vspace{2mm}\noindent\textbf{Primitives.} Les probabilités de
transition sont des fonctions des primitives suivantes. Soient deux
ensembles de n{\oe}uds $U$ et $W$, si chaque n{\oe}ud de $U$ peut infecter
chaque n{\oe}ud de $W$ avec probabilité $p$, alors la probabilité que $m$
n{\oe}uds de $W$ seront contaminés est~:
\begin{equation}
P_{cont}(m,p,|U|,|W|) = \textrm{pdf}_{\mathcal{B}}\left(m,1-(1-p)^{|U|},|W|\right)
\end{equation}
\noindent où $\textrm{pdf}_{\mathcal{B}}(k,p,n)$ est la densité de
probabilité d'une distribution binomiale de $n$ événements
indépendants avec probabilité $p$.

Les n{\oe}uds qui viennent d'être infectés peuvent contaminer la
destination au pas de temps suivant avec probabilité $\pi_\uparrow$,
tandis que des n{\oe}uds qui sont infectés depuis plus longtemps peuvent
le faire avec probabilité $1-q_i$ (un nouveau lien apparaît). Si $i$
n{\oe}uds sont infectés depuis plus de deux pas de temps et $j$ n{\oe}uds
viennent d'être infectés, alors la probabilité d'infecter la
destination au pas de temps suivant est~:
\begin{equation}
P_{succ}(i,j) = 1 - \pi_\downarrow^j q_i^i.
\label{p_succ}
\end{equation}

\vspace{2mm}\noindent\textbf{Probabilités de transition.} L'état
$Succ$ est absorbant. Toute transition depuis l'état $Init$ peut être
calculée comme une transition a partir de l'état $(0,1)$. Un état
$(i,j)$ peut aller ou bien dans l'état $Succ$ avec probabilité
$P_{succ}(i,j)$ ou bien dans un autre état $(i+j,j')$ avec probabilité~:
\begin{eqnarray}
  \lefteqn{\left( 1-P_{succ}(i,j) \right) \sum_{m=0}^{j'} \big( P_{cont}(m,\pi_\uparrow,j,N-1-i-j)} \nonumber \\
  & & \times P_{cont}(j'-m,1-q_i,i,N-1-i-j-m) \big).
\label{p_trans}
\end{eqnarray}

\vspace{2mm}\noindent\textbf{Taux de livraison.} Soit $\mathbf{T}$ la
matrice de transition de la chaîne de Markov, $\mathbf{i}$ le vecteur
d'état initial et $\mathbf{s}$ le vecteur d'état avec un coefficient
$1$ pour l'état $Succ$ et $0$ pour tous les autres. Ainsi, la
probabilité de succès, c.-à-d. d'être dans l'état $Succ$ en $d$ pas de
temps ou moins est~:
\begin{equation}
P_{deliv}( d, \alpha = 1 ) = \mathbf{i} \mathbf{T}^d \mathbf{s}.
\end{equation}

\subsection{Paquets plus petits que la capacité d'un lien \texorpdfstring{($\alpha < 1$)}{}}
%\subsection{Paquets plus petits que la capacité d'un lien ($\alpha < 1$)}
\label{subsec:p_smaller}

Lorsque la taille d'un paquet est plus petit que la capacité d'un
lien, ceux-ci peuvent effectuer jusqu'à $\lfloor \frac{1}{\alpha}
\rfloor$ sauts par pas de temps. Or nous avons supposé que la topologie du
réseau changeait instantanément au début de chaque pas de temps, avant
le \emph{premier} saut. Les sauts suivants se passent sur la
même topologie \emph{statique}. Pour en tenir compte, nous définissons
une matrice de propagation \emph{statique} $\mathbf{R}$ qui utilise
les même états que précédemment mais avec des probabilités de
transition légèrement modifiées. Sur une topologie statique aucun
nouveau lien n'apparaît donc $P_{succ}^{static}(i,j) = 1 -
\pi_\downarrow^j$ et la probabilité de transition d'un état $(i,j)$
vers $(i+j,j')$ devient~:
\begin{equation}
\left( 1-P_{succ}^{static}(i,j) \right) \cdot P_{cont}(j',\pi_\uparrow,j,N-1-i-j).
\end{equation}
Au final, la probabilité de succès du routage (c.-à-d. la probabilité
d'être dans l'état $Succ$ en moins de $d$ pas de temps) vaut~:
\begin{equation}
P_{deliv}(d,\alpha < 1) = \mathbf{i} \left( \mathbf{T} \mathbf{R}^{\lfloor \frac{1}{\alpha} \rfloor -1} \right)^d \mathbf{s}.
\end{equation}

\subsection{Paquets plus grands que la capacité d'un lien \texorpdfstring{($\alpha > 1$)}{}}
%\subsection{Paquets plus grands que la capacité d'un lien ($\alpha > 1$)}
\label{subsec:p_greater}

Les paquets plus grands que la capacité des liens ne peuvent utiliser
que ceux qui durent $\lceil \alpha \rceil$ ou plus pas de temps. Le
calcul exact de la probabilité de succès dans ce cas requiert de
distinguer les n{\oe}uds qui vont finir de recevoir une copie du paquet
dans $1,2,\cdots,\lceil \alpha \rceil$ pas de temps, ce qui devient
rapidement intractable. En revanche, il est possible de calculer
facilement des bornes supérieures et inférieures sur la probabilité de
succès en considérant des intervalles successifs et disjoints de
$\lceil \alpha \rceil$ pas de temps, et uniquement les liens qui
durent plus de $\lceil \alpha \rceil$ pas de temps. Ces derniers
seront dans la suite appelés les liens \emph{suffisament longs}.

La borne inférieure s'obtient en ne prenant en compte que les liens
\emph{suffisamment longs} qui existent ou apparaissent au \emph{début}
d'un intervalle. On ignore donc les liens qui apparaissent plus tard
dans l'intervalle, ce qui conduit à sous-estimer la propagation de
l'épidémie. Plus précisément, nous remplaçons $\pi_\uparrow$ par
$\pi_\uparrow q_c^{\lceil \alpha \rceil - 1}$ et $1-q_i$ par
$(1-q_i)q_c^{\lceil \alpha \rceil - 1}$ dans les formules \ref{p_succ}
et \ref{p_trans}.

La borne supérieure s'obtient en considérant que \emph{tout lien
  suffisamment long} qui apparaît pendant un intervalle pourra faire
passer la totalité du paquet à travers lui avant la fin de
l'intervalle. Par exemple, si $\alpha=2$ et qu'un lien
\emph{suffisamment long} apparaît au deuxième pas de temps de
l'intervalle, alors nous considérons que ce lien pourra transmettre la
totalité du paquet avant la fin de l'intervalle (c.-à-d. en $1$ pas de
temps). Ceci surestime évidemment le nombre de n{\oe}uds infectés à
chaque pas de temps. Plus précisément, nous remplaçons $\pi_\uparrow$
par $(\pi_\uparrow+\pi_\downarrow (1-q_i^{\lceil \alpha \rceil - 1}))
q_c^{\lceil \alpha \rceil - 1}$ et $1-q_i$ par $(1-q_i^{\lceil \alpha
  \rceil})q_c^{\lceil \alpha \rceil - 1}$ dans les formules
\ref{p_succ} et \ref{p_trans}.

Soient $\mathbf{T_{inf}}$ et $\mathbf{T_{sup}}$ les matrices de
transition obtenues pour les bornes inférieures et supérieures. La
probabilité d'acheminer le paquet en moins de $d$ pas de temps est
bornée par~:
\begin{equation}
\mathbf{i} \mathbf{T_{inf}}^{\frac{d}{\lceil \alpha \rceil}} \mathbf{s} \le P_{deliv}(d,\alpha > 1) \le \mathbf{i} \mathbf{T_{sup}}^{\frac{d}{\lceil \alpha \rceil}} \mathbf{s}.
\end{equation}

\subsection{Influence de la taille du paquet}
\begin{figure}[t]
  \centering
  \includegraphics{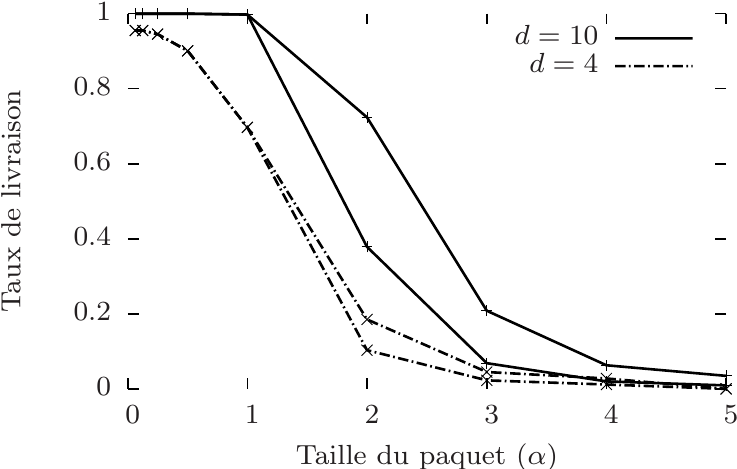}
  \caption{Influence de la taille du paquet sur le taux de livraison
    ($N=20$, $r=2.0$, $\lambda=10.0$) pour différentes valeurs du
    délai maximal toléré ($d$). Il y a deux lignes pour chaque valeur
    de $d$ qui correspondent aux bornes inférieure et supérieure.}
  \label{param_size}
\end{figure}

Les paquets de taille supérieure à la capacité du lien voient leur
taux de livraison se dégrader sérieusement, bien que ce soit à nuancer
pour de grands délais tolérés (fig.~\ref{param_size}). A l'inverse,
les paquets plus petits que la capacité du lien sont capables
d'effectuer plusieurs sauts par pas de temps. Ceci est
particulièrement avantageux lorsque les contraintes de délai sont
fortes ($d=4$ sur la fig.~\ref{param_size}), mais l'est moins lorsque
cette contrainte est relâchée. L'influence de la mobilité des n{\oe}uds
apparaît ici. En effet, puisque la taille des paquets est
proportionnelle au pas de temps $\tau$
(cf. section~\ref{subsec:hypotheses}), une mobilité importante
(c.-à-d. un $\tau$ petit) rend plus petite la capacité réelle des
liens (également proportionelle à $\tau$) et donc place davantage de
contraintes sur les tailles de paquet envisageables.

\subsection{Influence des autres paramètres}
\label{params}

\begin{figure}[t]
  \centering
  \subfloat[Nombre de n{\oe}uds \label{param_nodes}]{\includegraphics{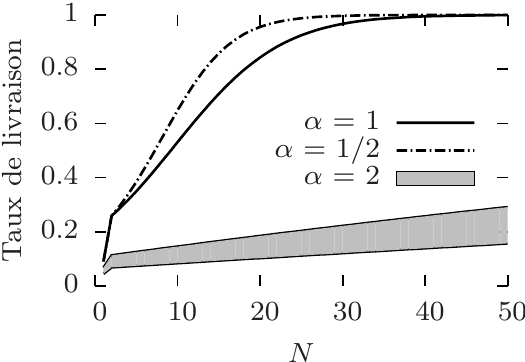}} \qquad
  \subfloat[Durée de vie moyenne des liens \label{param_life}]{\includegraphics{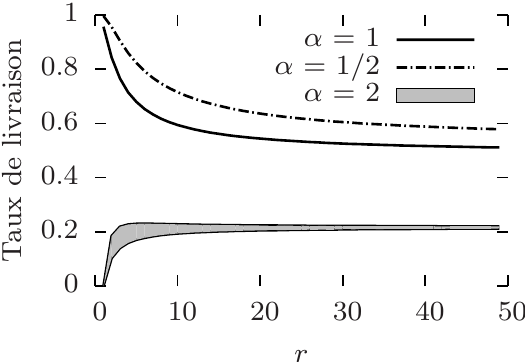}} \\
  \subfloat[Degré moyen des n{\oe}uds \label{param_degree}]{\includegraphics{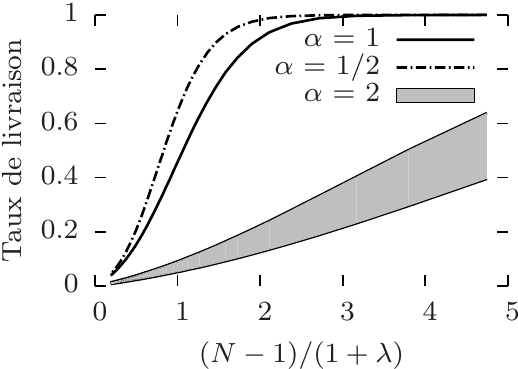}} \qquad
  \subfloat[Délai maximum \label{param_delay}]{\includegraphics{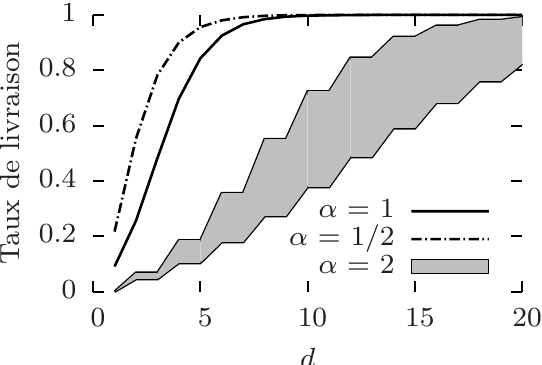}}
  \caption{Influence des paramètres du modèle sur le taux de
    livraison. Sauf mention contraire, $N=20$, $r=2.0$,
    $\lambda=10.0$ et $d=5$.}
  \label{params_influence}
\end{figure}

\noindent\textbf{Nombre de n{\oe}uds.}  (fig.~\ref{param_nodes}) Le taux
de livraison tend vers $1$ lorsque $N$ augmente. En effet, pour une
paire source/destination donnée, chaque nouveau n{\oe}ud est un nouveau
relais potentiel lors de la dissémination épidémique, ce qui ne peut
qu'aider le taux de livraison.

\noindent\textbf{Durée de vie moyenne des liens.}
(fig.~\ref{param_life}) Une plus courte durée de vie moyenne des liens
correspond à une topologie plus dynamique du réseau. En effet, de
petites valeurs de $r$ se traduisent par des temps de contact et
d'inter-contact plus courts (Eq. \ref{contact_inter_contact_times}) et
donc augmentent les opportunités de contact. De petits paquets
($\alpha \le 1$) peuvent en profiter et leur taux de livraison
augmente lorsque $r$ décroît. A l'inverse, une trop grande instabilité
des liens fait tendre le taux de livraison de paquets plus grands
($\alpha > 1$) vers $0$ car il existe alors moins de liens qui durent
plus d'un pas de temps.

\noindent\textbf{Degré moyen des n{\oe}uds.}  (fig.~\ref{param_degree})
Une plus grande connectivité accroît le taux de livraison. La forte
pente de la courbe lorsque $\alpha \le 1$ rappelle le phénomène de
percolation dans des graphes aléatoires réguliers quand le degré moyen
atteint $1$.

\noindent\textbf{Délai maximum.} (fig.~\ref{param_delay}) Toutes
choses égales par ailleurs, il y a clairement une valeur au-delà de
laquelle presque tous les paquets sont livrés. On peut relier ceci au
diamètre spatio-temporel de la topologie
sous-jacente~\cite{chaintreau_diam}.

\section{Résultats expérimentaux}
\label{sec:experimental}

Les résultats théoriques de la section précédente guident notre
intuition sur des scénarios réels. Bien que le modèle ne soit pas
quantitativement comparable à des résultats obtenus à partir de traces
de connectivité réelles à cause de propriétés ``petit-monde''
indésirables, il prédit précisément les relations entre le taux de
livraison, le délai maximum et la taille des paquets, comme nous le
verrons dans cette section.

\subsection{Jeux de données}
\label{datasets}

\begin{table}[t]
  \centering
\caption{Quelques traces Bluetooth.}
\begin{tabular}{lccc}
  %\hline
  & \textbf{MIT} & \textbf{Infocom} & \textbf{Rollernet} \\
  \hline
  \textbf{Durée (jours)}    & 365     & 3              & 0.125 \\
  \textbf{Période d'échantillonage (s)}& 600     & 120     & 15 \\
  \textbf{Nombre de capteurs}            & 100    & 41              & 62 \\
  \hline
\end{tabular}
  \label{blue}
\end{table}

La collecte de traces de connectivité sans-fil entre appareils
portables s'effectue typiquement à partir de sondages Bluetooth
périodiques. Dans ce papier, nous avons choisi d'étudier la trace
Rollernet~\cite{tournoux08_rollernet}, collectée lors d'une randonnée
roller, pour sa très courte période d'échantillonage. En effet, plus
la période d'échantillonage est longue,
plus il devient difficile de supposer qu'un contact équivaut à
l'existence d'un lien qui dure à peu près aussi longtemps que la
période d'échantillonage, ce qui est l'une de nos hypothèses de
base. C'est pourquoi, afin de pouvoir comparer théorie et pratique,
nous avons besoin de traces avec de très courtes périodes
d'échantillonage.

D'autres traces furent considérées, comme l'expérience ``Reality
Mining'' du MIT, dans laquelle chaque participant faisait tourner une
application sur son téléphone mobile qui relevait périodiquement sa
proximité avec les 100 autres participants pendant une année
complète~\cite{mit}. Le projet Haggle utilisa des Intel iMotes pour
mesurer les contacts entre participants de la conférence Infocom
2005~\cite{chaintreau_mobility}. Une rapide comparaison des ces traces
se trouve sur le tableau~\ref{blue}. La trace Rollernet est la seule à
proposer une période d'échantillonage suffisamment courte. D'une
certaine façon, on peut dire que les traces MIT et Infocom capturent
un sous-ensemble des opportunités de contact, tandis que la trace
Rollernet s'approche de l'évolution du graphe de connectivité.

\subsection{Méthodologie}

Nous étudions le taux de livraison réalisé par du routage épidémique
pour différentes tailles de paquets sur la trace de connectivité
sans-fil Rollernet~\cite{tournoux08_rollernet}.
Comme dans le modèle Markovien décrit à la section~\ref{sec:theory},
nous supposons que tous les liens ont la même capacité. La période
d'échantillonage dans Rollernet est de 15 secondes. La trace dure 3000
secondes. Toutes les 15 secondes pendant les premières 2000 secondes,
nous tirons au sort 60 couples source/destination pour une simulation
de routage épidémique.

\subsection{Des paquets plus petits augmentent le taux de livraison}

\begin{figure}[t]
  \centering
  \subfloat[Taux de livraison en fonction de la taille des paquets.\label{exp_results}]{\includegraphics{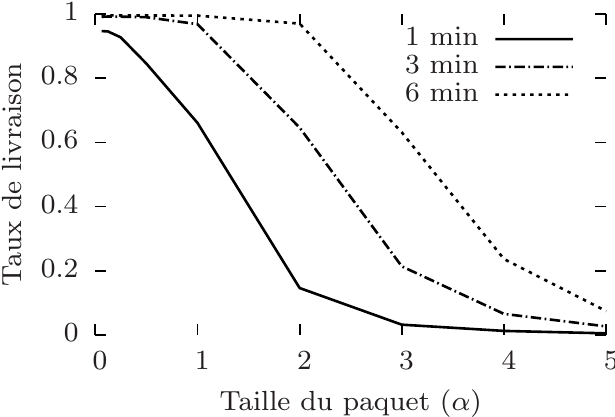}} \qquad
  \subfloat[Taille de paquet maximale capable de respecter un taux de livraison fixé en fonction du délai.\label{size_delay}]{\includegraphics{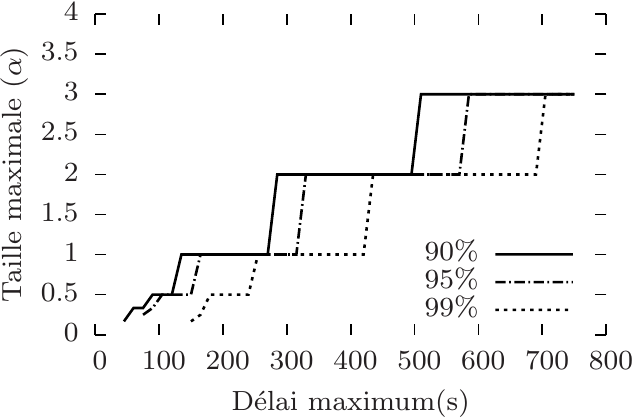}}
  \caption{Résultats sur la trace Rollernet.}
  \label{exp}
\end{figure}

Sur la fig.~\ref{exp_results}, le taux de livraison est constant et
proche de $1$ avant de chuter rapidement au-delà d'une certaine taille
de paquet qui dépend du délai fixé. Due a la mobilité très
élevée, la durée de vie moyenne d'un lien dans Rollernet est $26.2$
secondes et plus de la moitié des liens durent moins de $15$
secondes. Ainsi, des paquets de taille plus grande que $1$ perdent de
nombreuses opportunités de contact mais ceci peut être compensé par
une plus grande tolérance sur les délais. Ces observations
correspondent exactement aux résultats théoriques sur la taille, le
délai et la mobilité de la fig.~\ref{param_size}.

\subsection{Un gain borné pour des paquets plus petits}

Sur la fig.~\ref{exp_results}, lorsque le délai maximum est d'une
minute, le taux de livraison maximal est d'environ $0.95$. Prendre des
paquets plus petits n'y changera rien. Cette borne sur le gain obtenu
par de petits paquets apparaît car ils atteignent les limites de
performance du routage épidémique. En effet, la plus rapide diffusion
épidémique possible infectera, à chaque pas de temps, la totalité
d'une composante connexe dès lors que l'un de ses n{\oe}uds était
contaminé. Un paquet suffisamment petit pourrait produire cet effet,
mais des paquets encore plus petits n'auraient aucun avantage en
termes de taux livraison. Cette même limite de gain pour les petits
paquets est visible sur la courbe $d=4$ de la fig.~\ref{param_size}.

\subsection{Des délais courts nécessitent de petits paquets}

Afin de mieux comprendre la relation entre délai maximum et taille de
paquet convenable, la fig.~\ref{size_delay} trace la plus grande
taille de paquet qui est capable de respecter un taux de livraison
fixé, en fonction du délai maximum. Une contrainte forte sur le délai,
moins de quelques minutes par exemple, impose l'usage de petits
paquets afin d'obtenir un taux de livraison satisfaisant. \`A l'inverse,
le relâchement de cette contrainte de délai amène davantage de
flexibilité concernant la taille des paquets.

\section{Conclusion}

Dans ce papier, nous avons proposé un nouveau modèle de graphes
temporels aléatoires qui, pour la première fois, capture la
corrélation entre graphes de connectivité successifs. Les résultats
théoriques concernant l'interaction entre la mobilité des n{\oe}uds, le
délai maximum toléré et la taille des paquets sont confirmés
expérimentalement. En particulier, nous avons montré que, pour un
délai maximum fixé et une certaine mobilité des n{\oe}uds, la taille des
paquets a un impact majeur sur le taux de livraison. Ce résultat
devrait être pris en considération lors de la conception et
l'implémentation de nouveaux services mobiles.

\acknowledgements{Ce travail a été partiellement soutenu par le projet
  ANR Crowd ANR-08-VERS-006}

\bibliography{whitbeckcfip09}

\end{document}